\documentstyle[11pt,epsfig]{article}
\title{
Gluon Polarization In Nucleon}

\title{
Gluon Polarization In Nucleon}

\author{Abolfazl Shahveh$^{(a)}$, Fatemeh Taghavi-Shahri$^{(b)}$, and Firooz Arash$^{(a)}$\footnote{e-mail: farash@cic.aut.ac.ir}
\\
$^{(a)}$ Physics Department, Tafresh University, Tafresh, Iran \\
$^{(b)}$ School of Particles and Accelerators, Institute for
Research in Fundamental \\Sciences
(IPM) P.O. Box 19395-5531, Tehran, Iran\\
}
\date{\today}
\begin{document}
\maketitle
\begin{abstract}
In the context of the so-called valon model, we calculate
$\frac{\delta g}{g}$ and show that although it is small and
compatible with the measured values, the gluon contribution to the
spin of nucleon can be sizable. The smallness of $\frac{\delta
g}{g}$ in the measured kinematical region should not be
interpreted as $\delta g$ being small. In fact,  $\delta g$ itself
at small x, and  the first moment of the polarized gluon
distribution in the nucleon, $\Delta g(Q^2)$,  are large.

\end{abstract}
\section{INTRODUCTION}
The decomposition of nucleon spin in terms of its constituents have
been an active topic both from theoretical and experimental point of
views. It is established that the quark contribution, $\Delta
\Sigma$, to the nucleon spin is a small fraction of the nucleon
spin. Other sources that might contribute to the nucleon spin come
from gluon spin and the overall orbital angular momentum of the
partons. Thus, one can write the following spin sum rule for a
nucleon:
\begin{equation}
\frac{1}{2}=\frac{1}{2}\Delta \Sigma +\Delta g +L_{q,g}
\end{equation}
where $\Delta \Sigma$ is the quarks and anti-quarks contribution to
the nucleon spin, $\Delta g$ is the gluon contribution and $L_{q,g}$
represents the overall orbital angular momentum
contribution of the partons.\\
 In deep inelastic scattering, the gluon spin content of the
nucleon can be calculated from the $Q^2$ dependence of the
polarized structure function $g_1$. Experimentally, it is possible
to use semi-inclusive deep inelastic scattering processes to
measure $\frac{\delta g}{g}$ from helicity asymmetry in
photon-gluon fusion, $\gamma^* g\rightarrow q\bar{q}$ process. The
COMPASS collaboration \cite{1} have used this method and find a
rather small value for $\frac{\delta g}{g}=0.024\pm 0.080 \pm
0.057$. The smallness of $\frac{\delta g}{g}$ cannot by itself
rule out the possibility of a large value for the first moment,
$\Delta g$, of the gluon polarization. In fact, when the singlet
axial matrix element $a_0$ was found to be much smaller than the
contribution expected from quark-parton model, it was suggested
that the deference could be accounted for by a large contribution
from the gluon spin:
$\Delta\Sigma=a_{0}-N_{f}\frac{\alpha_s}{2\pi}\Delta g$. This
would require a value of $\Delta g \sim 3$ at $Q^{2}=3$ $GeV^{2}$
in order to obtain the expected value of $\Delta \Sigma$.
Moreover, Altarelli and Ross \cite{2} and Effremov et.al \cite{3}
have shown that polarized gluon makes a scaling contribution to
the first moment of the polarized structure function, $g_{1}$,
which means that it must be large at higher
momentum scales. \\
The total quark spin contribution $\Delta \Sigma$ to the nucleon
spin is fairly well determined and gives a value around 0.4. There
is no direct determination of orbital angular momentum component,
and one is not expected in the near future. In contrast to $\Delta
\Sigma$, knowledge about gluon polarization is limited. The existing
and the emerging data on $\frac{\delta g(x,Q^2)}{g(x,Q^2)}$ cannot
rule out the negative and/or zero polarization for gluon, including
a possible sign change. There are mainly three method to access
gluon polarization: (1) polarized deep inelastic scattering, in
which one would parameterize quark and gluon densities and fit them
to the data on polarized structure function $g_{1}(x, Q^{2})$. Gluon
enters into the analysis through the $Q^2$ evolution, but the
limited range of $Q^2$ leads to not so precise determination of
$\delta g(x)$. Recent data suggests that global fits with positive,
negative, zero, and sign changing $\delta g(x)$ provide equally good
agreement. (2) Using $c \bar{c}$ production in semi-inclusive deep
inelastic processes by $\gamma-g$ fusion. (3) via single particle
production in polarized p-p
collision.\\
In this paper we determine the gluon polarization in the polarized
proton using the so called ${\it{valon}}$ model,  as described bellow.\\
\section{ The valon model description of nucleon}
Deep inelastic scattering reveals that the nucleon has a
complicated internal structure. Other strongly interacting
particles also exhibit similar structure. However, under certain
conditions, hadrons behave as consisting of three (or two)
constituents. Therefore, it seems to make sense to decompose a
nucleon into three constituent quarks called U and D. We identify
them as $\it{valons}$. A valon has its own internal structure,
consisting of a valence quark and a host of $q \bar{q}$ pairs and
gluons. The structure of a valon emerges from the dressing of a
valence quark with $q \bar{q}$ pairs and gluons in perturbative
QCD. We take the view that when a nucleon is probed with high
$Q^{2}$ it is the internal structure of the valon that is
resolved. The valon concept was first developed by R. C. Hwa \cite
{4}, and in Refs. \cite {5,6,7} it was utilized to calculate
unpolarized structure functions of a number of hadrons. This
representation is also used to calculate the polarized
structure of nucleon. The details can be found in \cite {8,9} . \\
We have worked in $\overline{MS}$ scheme with $\Lambda_{QCD}=0.22$
GeV and $Q_{0}^{2}=0.283$ $GeV^2$. The polarized and unpolarized
structure of a valon is calculated in the framework of
Next-to-Leading order in QCD. Then, the polarized (unpolarized)
structure function of the nucleon is obtained by the convolution of
the valon structure with the valon distribution in the hosting
nucleon:
\begin{equation}
g^{h}_{1}(x,Q^{2})=\sum_{\it{valon}}\int_{x}^{1}\frac{dy}{y}
\delta G_{\it{valon}}^{h}(y) g^{\it{valon}}_{1}(\frac{x}{y},Q^{2})
\end{equation}
where $\delta G_{\it{valon}}^{h}(y)$ is the helicity distribution
of the valon in the hosting hadron and
$g^{\it{valon}}_{1}(\frac{x}{y}, Q^{2})$ is the polarized
structure function of the valon. A similar relation can also be
written for the unpolarized structure function, $F_{2}$. We
maintain the results of Ref. \cite{8} for the polarized structure
function, but re-analyze the unpolarized case. This is necessary
in order to arrive at a consistent conclusion on $\frac{\delta
g}{g}$. In the moment space the initial densities for both
polarized and unpolarized densities  of the partons in a valon are
taken to be as follows,
\begin{equation}
\left(\begin{array}{c} \delta q^n(Q_{0}^{2})
\\\delta g^n(Q_{0}^{2})
\end{array}\right)=\left(\begin{array}{c} q^n(Q_{0}^{2})
\\g^n(Q_{0}^{2})
\end{array}\right)=\left(\begin{array}{c} 1
\\0 \end{array}\right)
\end{equation}
The above initial densities means that if $Q^{2}$ is small enough,
at some point we may identify
$g^{\it{valon}}_{1}(\frac{x}{y},Q^{2})$ and
$f^{\it{valon}}_{2}(\frac{x}{y},Q^{2})$ as $\delta (z-1)$, for the
reason that we cannot resolve its internal structure at such
$Q^{2}$ value. Here $f^{\it{valon}}_{2}(\frac{x}{y},Q^{2})$ is the
unpolarized structure function of the valon. \\
In figure (1) a sample of results for the unpolarized structure
function, $F_{2}$, is presented. The data points are from \cite
{10,11}. Similar results are also obtained at different kinematics \cite{5}.\\
\begin{figure}
\centerline{\begin{tabular}{cc}
\epsfig{figure=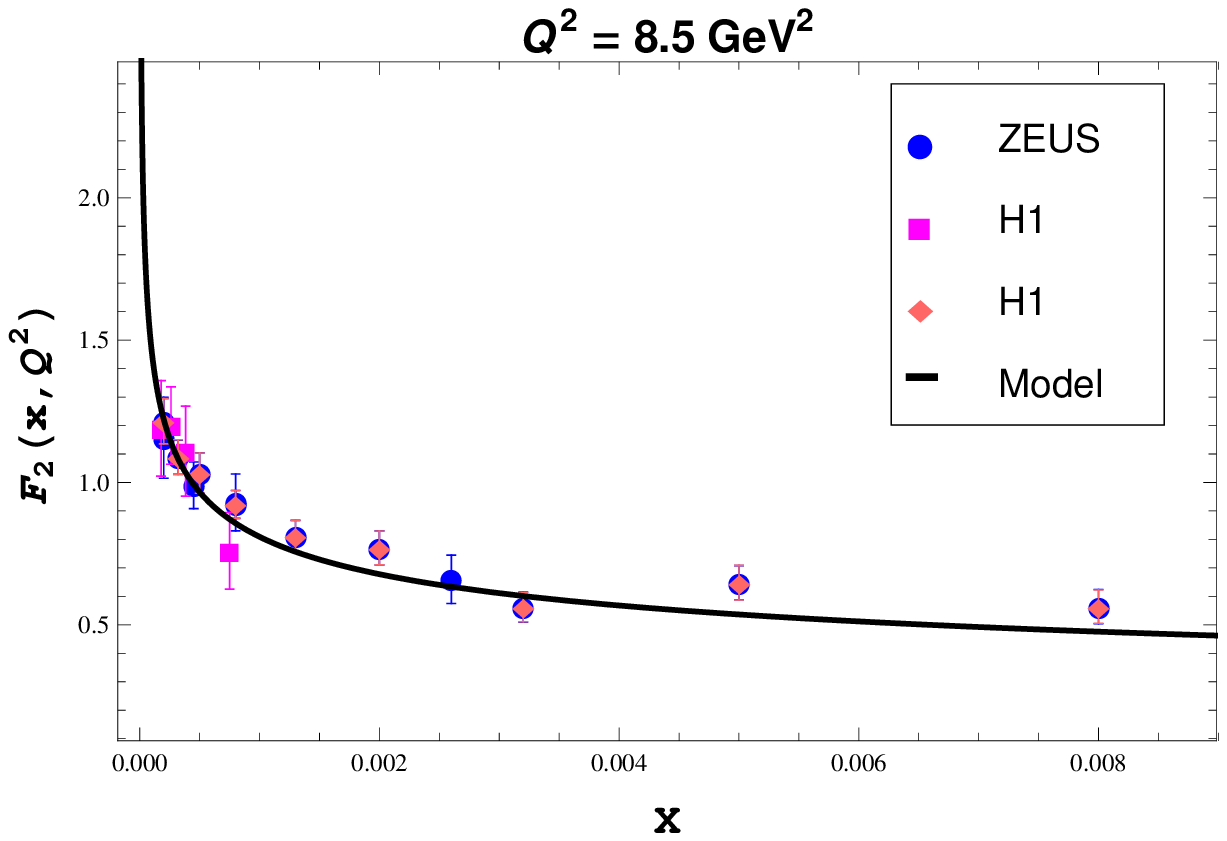,width=7cm}
\epsfig{figure=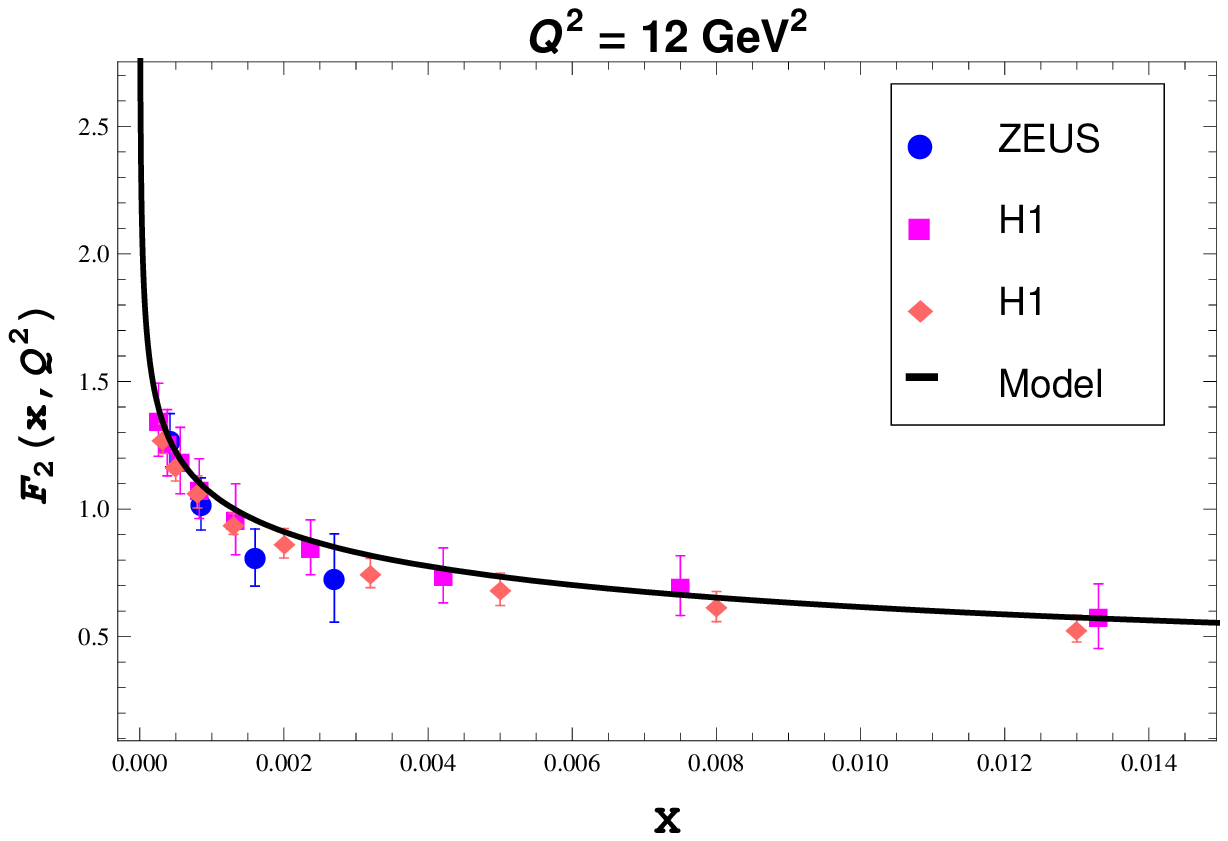,width=7cm}
 \end{tabular}}
\caption{\footnotesize Unpolarized structure function of proton ,
 $F_{2}^{p}$ at $Q^2=8.5$ and $12$ $GeV^2$. Data points are from \cite
{10,11} . } \label{figure 1.}
\end{figure}
Figure (2) shows the polarized structure function of proton,
$xg_{1}^{p}$, obtained from the model, along with the available
data from various experiments \cite{12,13,14,15,16,17}. It is
important to note that our analysis does not rely on any kind of
data fitting. The structure of a valon is obtained simply from QCD
processes via DGLAP evolution \cite {18,19,20}.\\
\begin{figure}
\begin{center}
\epsfig{figure=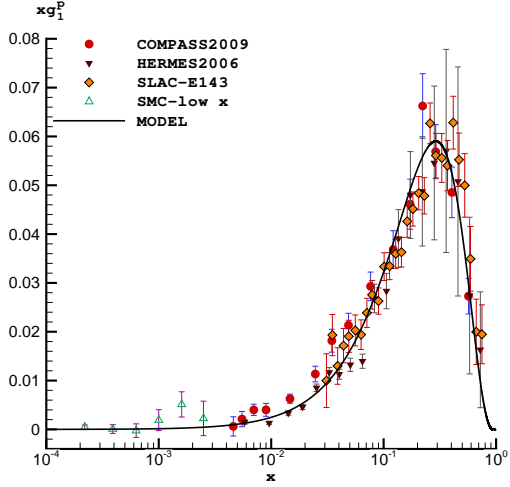,width=8cm}
 \caption{\footnotesize
Polarized structure function of proton ,
 $xg_{1}^{p}$, as a function of $x$. Solid curve is the model results at $Q^2=5 GeV^2$.
The data points are from Refs. [12-17].} \label{figure 2.}
\end{center}
\end{figure}
 Our main concern here is to determine the polarized and unpolarized
gluon distributions and hence, the ratio $\frac{\delta
g(x,Q^2)}{g(x,Q^2)}$. The gluon is a component of the singlet
sector of the evolution kernel. Their moments are given as
\begin{equation}
 \left( \begin{array}{c}
 \delta M_{S}(n,Q^{2}) \\ \delta M_{G}(n,Q^{2})
 \end{array} \right)
=\{{\bf{L}}^{-(\frac{2}{\beta_{0}})\delta \hat{P}^{(0)n}}+
\frac{\alpha_{s}(Q^2)}{2 \pi}{\bf{\hat{U}
L}}^{-(\frac{2}{\beta_{0}})\delta
\hat{P}^{(0)n}}-\frac{\alpha_{s}(Q_{0}^{2})}{2
\pi}L^{-(\frac{2}{\beta_{0}})\delta
\hat{P}^{(0)n}}{\bf{\hat{U}}}\}\left( \begin{array}{c}
 {1} \\{0}
 \end{array}\right)
\end{equation}
where ${\bf{L}} \equiv \alpha_{s}(Q^2)/\alpha_{s}(Q^{2}_{0})$, and
$\delta \hat{P}^{(0)n}$ is $2\times 2$ singlet matrix of splitting
functions, given by
\begin{eqnarray}
\delta \hat{P}^{(0)n}= \left ( \begin{array} {c} \delta
P^{(0)n}_{qq} \hspace{0.75cm} 2f\delta P^{(0)n}_{qg} \\ \delta
P^{(0)n}_{gq}\hspace{0.75cm}\delta P^{(0)n}_{gg}
\end{array} \right ),
\end{eqnarray}
$\delta P^{(0)n}_{lm}$ are the $n^{th}$ moments of the polarized
splitting functions and ${\bf{U}}$ accounts for the 2-loop
contributions as an extension to the leading order. The explicit
forms of these functions are given in \cite {21} in the
next-to-leading order. Now it is straightforward to calculate the
moments of the polarized partons inside a valon at any $Q^{2}$
value. They are given in Ref. \cite{8}. The densities are obtained
by an usual inverse Mellin transformation. To obtain the polarized
parton distributions in a hadron, one needs to convolute the
results with the valon distribution in the hadron. In figure (3)
the first moments of the polarized partons in the proton are shown
as a function of $Q^2$.
\begin{figure}
\begin{center}
\epsfig{figure=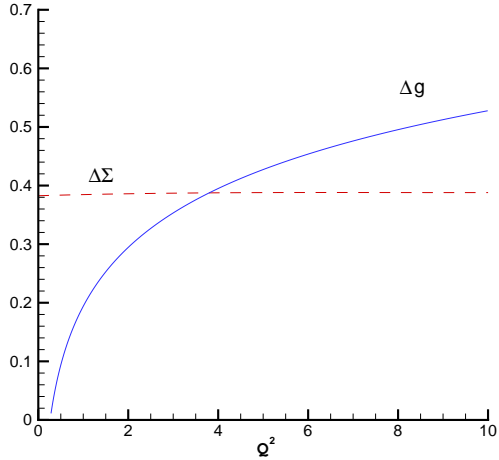,width=8cm}
 \caption{\footnotesize
First moments of polarized quark, $\Delta\Sigma$, and gluon,
$\Delta g$, in the proton as a function of $Q^2$. } \label{figure
3.}
\end{center}
\end{figure}
In spite of the fact that we have started with $\Delta g=0$ at the
starting scale, it grows rapidly with increasing $Q^2$. This
behavior of gluon polarization can be related to the positive sign
of the pertinent anomalous dimension $\delta \gamma^{(0)1}_{qg}$.
The positivity of the anomalous dimension dictates that the
polarized quark preferably to radiate a gluon with helicity
parallel to the quark polarization. Since the net quark spin in a
valon is positive, it follows that perturbatively radiated gluon
from quarks must have $\Delta g>0$. We also note that the growth
rate of $\Delta g$ is especially fast for the relatively low
$Q^2$. In order to satisfy the sum rule in Equation (1) it
requires that the orbital angular momentum component to be
negative and decreasing as $Q^2$ increases \cite {9}.\\
Figures (1) and (2) demonstrate that the model can accommodate the
experimental data on structure functions fairly accurately. The
calculated polarized gluon distributions, $x \delta g(x,Q^2)$, are
shown in figure (4) as a function of $x$ for several values of
$Q^2$. We have also shown the results at $Q^2=5$ $GeV^2$ and
compared it with the global fits from \cite {22,23,24}.
\begin{figure}
\centerline{\begin{tabular}{cc}
 \epsfig{figure=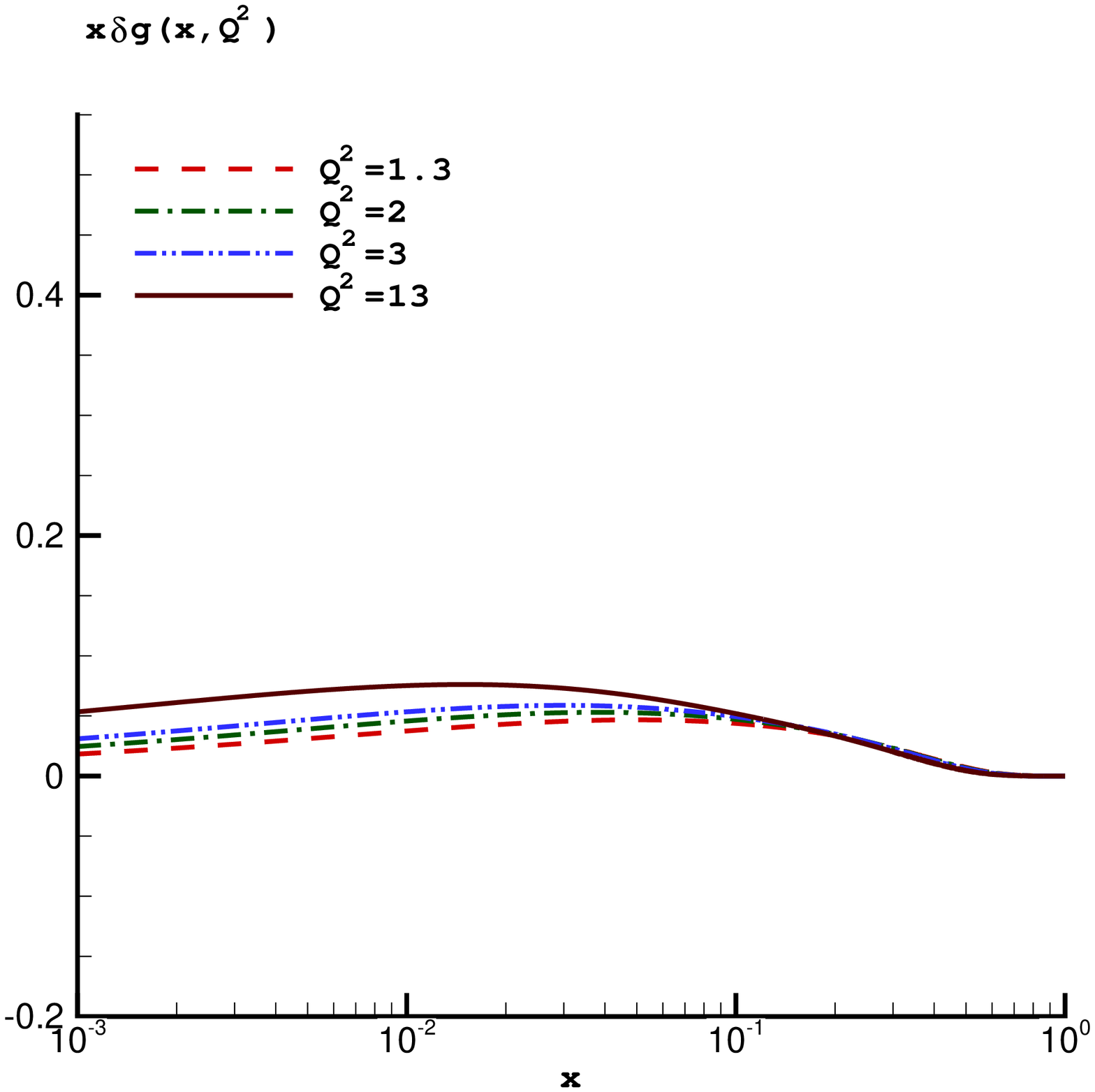,width=8cm}
\epsfig{figure=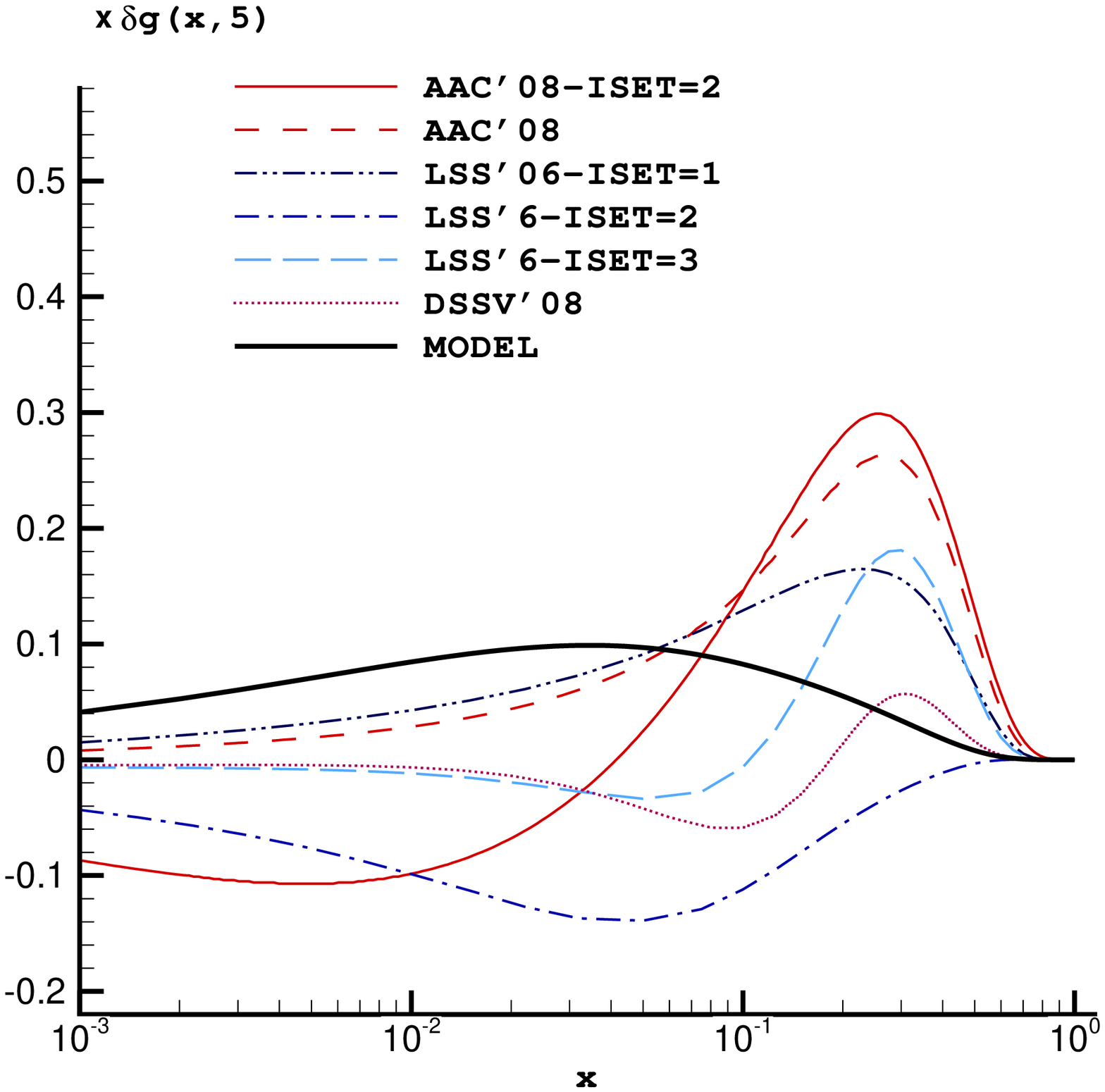,width=8cm}
\end{tabular}}
\caption{\footnotesize  {\it{Left:}} Polarized gluon distribution
function, $x \delta g(x,Q^2)$, for some values of $Q^2$.
{\it{Right:}} Comparison of the results of the present model and
global fits from Ref. \cite {22,23,24} at a single value of $Q^2=5
GeV^2$. } \label{figure 4.}
\end{figure}
The unpolarized gluon distribution is also shown in figure (5) and
is compared with the results obtained from
various global fits \cite {25,26,27,28} .\\
 \begin{figure}
\centerline{\begin{tabular}{cc}
 \epsfig{figure=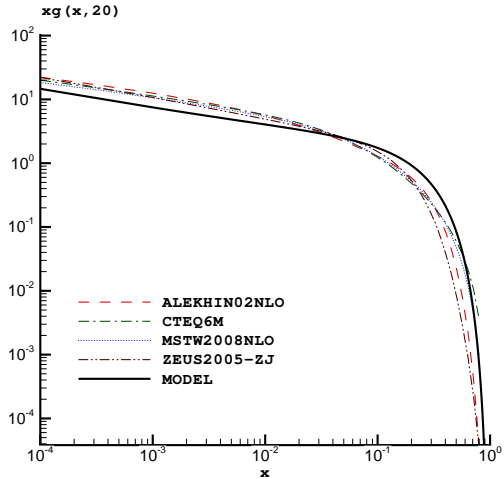,width=8cm}
\end{tabular}}
\caption{\footnotesize The model result for the unpolarized gluon
distribution, $xg(x)$, at $Q^2=20  GeV^2$. Also shown are the
results from the global fits. } \label{figure 5.}
\end{figure}
Our results for the sea quark polarization is consistent with
zero, and yields a positive value for the first moment of gluon,
$\Delta g(Q^2)$,  which increases with $Q^2$ reaching a value of
around 0.5
at $Q^{2}=10$ $GeV^2$ as can be seen from figure (3).\\
It is now straight forward to calculate the ratio $\frac{\delta
g(x)}{g(x)}$ in the proton. This ratio is calculated and shown in
figure (6). The calculation is done for the proton at each value
of $Q^2$ corresponding to the experimental kinematics. This allows
us to make a meaningful comparison of our results with the
experimental data. The apparent wide band in the figure is
actually seven closely packed curves corresponding to the seven
individual values of $Q^2$s at which the data are measured.
Apparently, HERMES high $p_T$ (2000) \cite {29} and COMPASS open
charm  data disagree with our results. However, these two data
points are the least accurate ones with very large error bars. In
contrast, our results are in good agreement with the remaining
experimental points, including the very recent one from HERMES and
COMPASS \cite {29,30,31,32,33,34,35}.
\begin{figure}
\begin{center}
\epsfig{figure=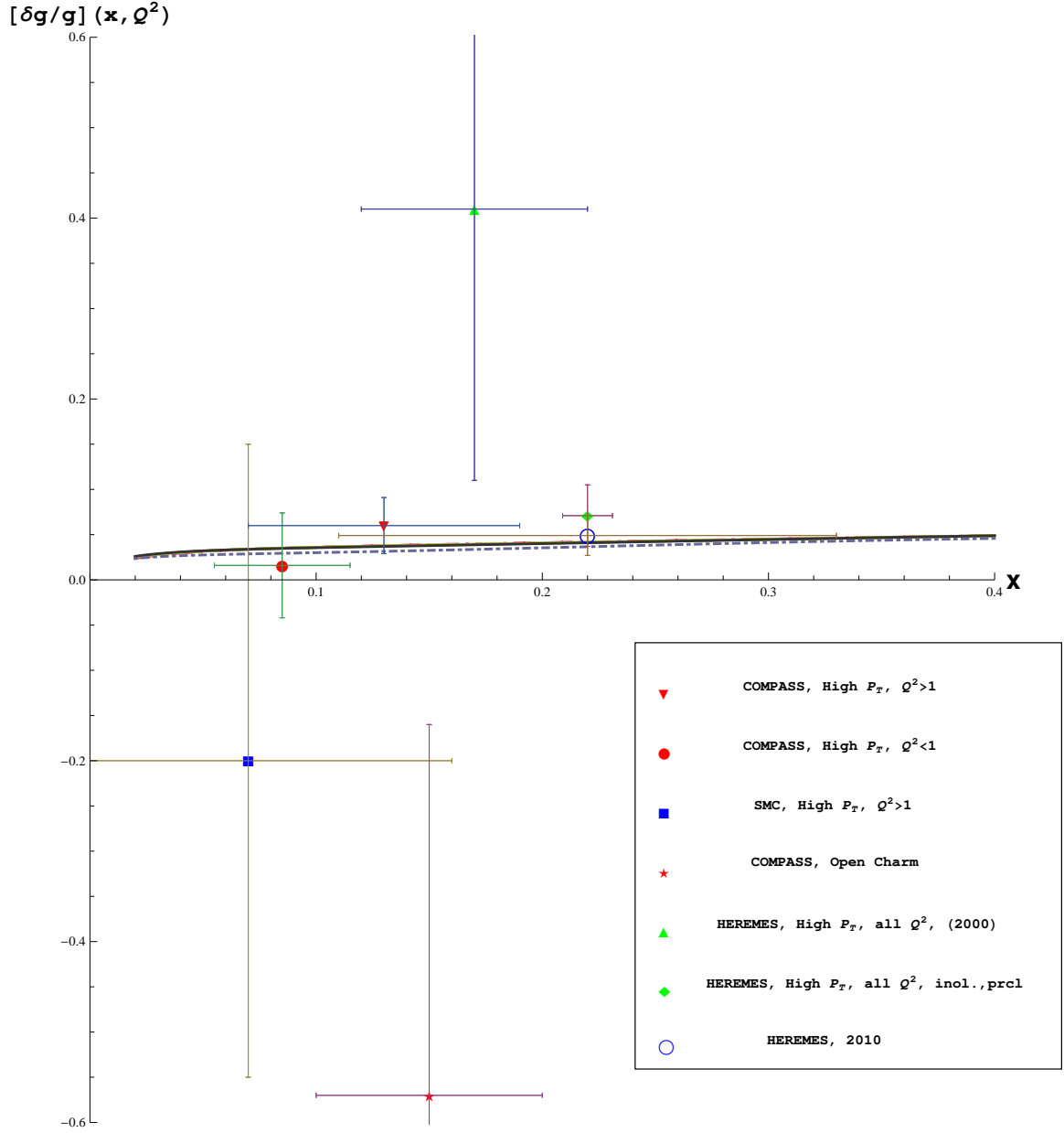,width=16cm}
 \caption{\footnotesize
The ratio $\frac{\delta g(x,Q^2)}{g(x,Q^2)}$ calculated in the
valon model and compared with the exist experimental data. The
apparent wide band in the figure is actually seven closely packed
curves corresponding to the seven values of $Q^2$s at which the
data points are measured . The data are from Refs. \cite
{29,30,31,32,33,34,35}. } \label{figure 6.}
\end{center}
\end{figure}

\section{Conclusion}
We calculated gluon polarization in a polarized proton in the
valon model and compared it with the existing data, including the
most recent one from HERMES collaboration \cite {30}. Since the
experimental data are obtained at different $Q^2$ values, the
calculations are also carried out at the corresponding $Q^2$,
individually. It is evident from the results that the polarized
valon model of nucleon not only agrees with the existing data on
$g_1$ but also provides a clear resolution for the spin problem.
We maintain the view that $\Delta g(Q^2)$ is positive and
increases with $Q^2$. The growth of $\Delta g(Q^2)$ in part is
compensated by a negative and large orbital angular momentum,
$L_{q,g}$. Although, we have not calculated $L_q$ and $L_g$
individually, but the overall $L_{q,g}$ is given in \cite {8}.\\
This suggests that even if $\delta g(x,Q^2)$  maybe small at
relatively large x, but the first moment of gluon polarization in
the proton is  sizable. In fact, $\delta g(x,Q^2)$ is quiet large
at small x.

\section*{Acknowledgments}
 We would like to thank Professor Mauro Anselmino and Professor  Jacques Soffer
for their critical reading of the manuscript and  for their useful
suggestions.

\end{document}